# Schrödinger's field equation

View Online   Export Citation   CrossMark

Jacek Mroczkowski[a])

**AFFILIATIONS**
7 Pickman Drive, Bedford, Massachusetts 01730,

[a)]Author to whom correspondence should be addressed: JAM000705@gmail.com

**ABSTRACT**

The *intrinsic* and *dynamic* kinetic energies and the potential energies of electron states in the hydrogen atom were determined using the operator formalism in Schrödinger's nonrelativistic equation. Intrinsic energies were determined using the momentum operator, while for $\ell \neq 0$, the additional *dynamic* energies of the spinning fields were determined using the angular momentum operator. All 10 states up to the principal quantum number n = 3 and all 4 m states of n = 7, $l$ = 3 were analyzed. The two forms of kinetic energy can only be explained with an electron *field* representation. All total kinetic and potential energies conformed with the well-known $1/n^2$ rule. Angular momentum analysis of the $2P_{1/2}$ state provided a field spinning rate; in addition, the dynamic kinetic energy of the spinning field determined by both operator analysis and explicit calculation based on the spinning rate gave the same energy results.

## I. INTRODUCTION

Most physicists now accept that particles are really *particle fields*. However, there is a problem whenever particles are described or implied to be *point structures*. This happens often in nonrelativistic quantum mechanics (QM) books with figures showing electrons orbiting nuclei like particles,[1–3] and it is especially misleading for physics students. One QM course book at MIT is replete with the term "particle" while stating "*the particle has to be somewhere.*"[4] Even when wave packets are briefly introduced later in this book, the term *field*, as in *particle field*, or *matter field* is never used. A recent edition of this book states, "*The electron is a structure-less particle.*"[5] A similar point particle description, with no mention of fields, appears in another QM book.[6] Both books are excellent, except for not explaining the true *field* nature of all particles. To quote physicist Roland Omnes,[7] "*The most important consequence of the Uncertainty relations for interpretations is their incompatibility with an intuitive representation of a particle being a point in space.*"

Relativistic quantum mechanics makes it clear that particles are really *particle fields*. Carlo Rovelli, in his book "*Reality is Not What it Seems,*" devotes a whole section entitled "*Fields and particles are the same thing.*"[8] Nobel laureate Frank Wilczek writes, "There is no need to introduce two different *sorts of fundamental ingredients, fields and particles, after all. Fields rule.*"[9] Nobel laureate Steven Weinberg states the same: "*Particles are just bundles of field energy.*"[10] Art Hobson puts it more bluntly: "*There are no particles, there are only Fields.*"[11] Hobson then goes further, describing Schrödinger's non-relativistic equation as a *field* equation.[12] This study supports that observation using different arguments based on the hydrogen (H) atom.

Wave functions and fields are the same entities, but the meaning of $\Psi^*\Psi$ for wave functions is occasionally incorrectly associated with a particle's position probability. When instead, $\Psi$ is recognized as a field, it is clear that it is the *whole field* that can transition to another *whole field* state.

The separation of variables in Schrödinger's equation for the H atom does not alone reveal if the electron is a particle or field. However, by using the energy operators in the equation on the wave functions (resulting from the separation of variables), both electron *intrinsic* and *dynamic* kinetic energies can be determined. These two distinct energies will provide a basis for identifying the electron in the H atom as a field.

Erwin Schrödinger introduced his wave equation in 1926, insisting that the electron charge was distributed as $-e\Psi^*\Psi$, not the probability of finding a particle electron.[13(a)–13(c),14(a)] Einstein supported this view.[14(b),15] Wigner later argued that electrons cannot be particles based on "Causality and invariance."[16] Schrödinger's biographers (13, 14) maintain that throughout his life, he defended the wave or field interpretation of his equation. However, at the fifth

Solvay gathering of physicists in 1927, Bohr dominated the conference and forced a Copenhagen interpretation of wave mechanics based on real particles.[14(c),15] This immediately gave an attractive computational "package," which worked,[14(d)] but the particle interpretation was just a *guess*. The interpretational differences were still on display more than 30 years later when Max Born, a real particle advocate, had to be the substitute speaker for Schrödinger.[17]

The choice of representation may have been different if physicists Bernard d'Espagnat, Roland Omnes, and John Bell had been around to share their views at the 1927 conference. D'Espagnat:[18] "*if the best we can do to describe reality is to resort at the same time to two contradictory pictures, then quite obviously we cannot claim to describe reality as it really is.*" Omnes:[19] "*Something real cannot be a bird and a tree, a stone and a sound, a wave and a particle.*" Bell:[20] "*that the idea that an electron in a ground state hydrogen atom is as big as the atom is perfectly tolerable—and may even be attractive.*"

Two reasons for the interpretational disconnect are suggested. First, as Hobson puts it, "*part of the problem is a lack of communication between low* energy physicists and high energy physicists who characterize everything with *fields.*"[21] Ironically, part of the problem is the pervasive use of the term *particle* with no qualifiers, although the real meaning is usually understood as that of a *particle field*. Repetitive use of words can bias beliefs. Second, in an analysis of the state of physics, physicist Lee Smolin identifies two problems in academia: frequent university *resistance to change* and *risk aversion*; both are cited many times in the later part of his book *The Trouble with Physics.*[22] Smolin argues that risk and imagination have often been key to major progress, but in academia they are less supported than conservative research. Indeed, scientists still on a career path might not wish to be sidelined into any controversial issue.

Louis de Broglie suggested that at a fundamental level matter exhibited wave properties ($\lambda = h/p$) after contemplating if *moving* particles could be better represented by waves. Schrödinger independently came up with a nonrelativistic Eq. (1) for *bound states* in which V, E are the potential and total energy,

$$-\frac{\hbar^2}{2m}\nabla^2\psi + V\psi = E\psi. \qquad (1)$$

The interpretational differences in Schrödinger's equation appear to be partly based on two misunderstandings. First, the electron's potential energy term in many problems appears as $V = (-e^2/r)$. This could suggest a *particle* instead of a *charged field*—$e\Psi^*\Psi$, where the attributes of charge and mass are *already proportionally distributed* throughout the field space,[23(a),24] as will be confirmed in the following H atom study. Second, although the velocity of particles is given by their momentum/mass ratio, as Richard Feynman put it, there are two kinds of momenta.[25] The first pertains to *bound* momentum k-states like that of the "particle in a box" problem, when the momenta of the forward states $e^{ikx}$ are canceled by those of the backward states $e^{-ikx}$. The second, considered by de Broglie, is for an unconstrained *moving* "particle."

Schrödinger's wave functions or fields in the H atom are *volume entities* with *inversion symmetry*. Nothing in his equation suggests any particle electron radial movement to satisfy the so-called probability function. The Laplacian is "designed" to operate on fields; spatial derivative operations on point particles are meaningless. Other applications of the equation are qualitatively evaluated, showing why the particle representation is unable to satisfy certain criteria.

## II. ANALYSIS

H atom states have half integral total angular momentum J because of intrinsic spin, but it is not necessary to use a relativistically compliant equation to determine electron energies to first order. All m states in each n, J, *l* subshell will be confirmed to have the *same intrinsic* and *dynamic* energies. Changing from an *l*, m basis to a J, $J_z$ basis *merely mixes equal energy m states while adding spin*. Since the sums of the two (coupling coefficients)$^2$ in each $|J, J_z\rangle$ state are unity, all $|J, J_z\rangle$ states with the same n, J, *l* will have both the same field energies and also the same very small spin orbit energy offsets,[26] which are irrelevant to this study. With electron field velocities to be shown to be ~0.3% of c, nonrelativistic energy analysis of the *l*, m states to first order is valid, with relativistic analysis reducing to the Schrödinger limit.[27]

While it may be tempting to differentiate between the two forms of kinetic energy as *Static* and *Dynamic,* the term *Static* is not appropriate because the field does pulsate. If the center of mass pulsates, a reduced mass m must be used.[28] The term *intrinsic* has therefore been chosen for the energy component *not* associated with the additional energy in the spinning field.

### A. Electron energies in hydrogen

The two kinetic energy operators[29] are

$$-\frac{\hbar^2}{2m}\nabla^2 = \underbrace{\frac{\mathbf{P}_r^2}{2m}}_{\text{Intrinsic energy operator}} + \underbrace{\frac{\mathbf{L}^2}{2mr^2}}_{\text{Dynamic energy operator}}. \qquad (2)$$

Substituting for the operators[30] and multiplying with $\Psi^*$ from the left gives

$$-\frac{\hbar^2}{2m}\Psi^*\left[\frac{1}{r^2}\frac{\partial}{\partial r}\left(r^2\frac{\partial\Psi}{\partial r}\right) + \frac{1}{r^2\sin\theta}\frac{\partial}{\partial\theta}\left(\sin\theta\frac{\partial\Psi}{\partial\theta}\right)\right.$$
$$\left. + \frac{1}{r^2\sin^2\theta}\frac{\partial^2\Psi}{\partial\Phi^2}\right] + \Psi^*V\Psi = E\Psi^*\Psi, \qquad (3)$$

where V and E were defined in Eq. (1). $\Psi$ can be expressed as the product of radial and angular $\theta$, $\varphi$ components, i.e., $\Psi = C_r R_{n,l}(r) \times \Theta(\theta) \times \vartheta(\varphi)$ in Table I, where $C_r$ is the radial normalization constant. The derivation of the radial function for n = 7 and *l* = 3 and the four m state energy analyses are given in the supplementary data.[31] The potential energy $V_{n,l}$ of the fields is

$$V_{nl} = eC_r^2\int_{\text{vol}}\frac{1}{r} - \underbrace{e|\Psi(r,\theta,\phi)|^2 r^2\sin\theta dr d\theta d\phi}_{\text{Electron field charge }\delta\rho(r,\theta)\text{ in volume dV}}$$
$$= -e^2 C_r^2\int_o^\infty rR_{nl}(r)^2 dr. \qquad (4)$$

**TABLE I.** Three normalized eigenfunctions of each state. $a = \hbar^2/me^2$.

| n, l, m | $C_r^2$ | R(r) | Polar $\Theta(\theta)$ | Azimuth $\vartheta(\varphi)$ | $n\ell_j$ | No. of $J_z$ |
|---|---|---|---|---|---|---|
| 1, 0, 0 | $\frac{4}{a^3}$ | $e^{-r/a}$ | $\frac{1}{\sqrt{2}}$ | $\frac{1}{\sqrt{2\pi}}$ | $1\,S_{1/2}$ | 2 |
| 2, 0, 0 | $\frac{1}{8a^3}$ | $(2 - \frac{r}{a})e^{-r/2a}$ | $\frac{1}{\sqrt{2}}$ | $\frac{1}{\sqrt{2\pi}}$ | $2\,S_{1/2}$ | 2 |
| 3, 0, 0 | $\frac{4}{27a^3}$ | $\left(1 - \frac{2r}{3a} + \frac{2r^2}{27a^2}\right)e^{-r/3a}$ | $\frac{1}{\sqrt{2}}$ | $\frac{1}{\sqrt{2\pi}}$ | $3\,S_{1/2}$ | 2 |
| 2, 1, 0 | $\frac{1}{24a^5}$ | $re^{-r/2a}$ | $\sqrt{\frac{3}{2}}\cos\theta$ | $\frac{1}{\sqrt{2\pi}}$ | $2\,P_{1/2}$ | 2 |
| 2, 1, ±1 | $\frac{1}{24a^5}$ | $re^{-r/2a}$ | $\frac{\sqrt{3}}{2}\sin\theta$ | $\frac{e^{\pm i\phi}}{\sqrt{2\pi}}$ | $2\,P_{3/2}$ | 4 |
| 3, 1, 0 | $\frac{32}{3^7 \cdot a^3}$ | $\left(\frac{r}{a} - \frac{r^2}{6a^2}\right)e^{-r/3a}$ | $\sqrt{\frac{3}{2}}\cos\theta$ | $\frac{1}{\sqrt{2\pi}}$ | $3\,P_{1/2}$ | 2 |
| 3, 1, ±1 | $\frac{32}{3^7 \cdot a^3}$ | $\left(\frac{r}{a} - \frac{r^2}{6a^2}\right)e^{-r/3a}$ | $\frac{\sqrt{3}}{2}\sin\theta$ | $\frac{e^{\pm i\phi}}{\sqrt{2\pi}}$ | $3\,P_{3/2}$ | 4 |
| 3, 2, 0 | $\frac{8}{5 \cdot 3^9 \cdot a^7}$ | $r^2 e^{-r/3a}$ | $\sqrt{\frac{5}{8}}(3\cos\theta^2 - 1)$ | $\frac{1}{\sqrt{2\pi}}$ | $3\,D_{3/2}$ | 4 |
| 3, 2, ±1 | $\frac{8}{5 \cdot 3^9 \cdot a^7}$ | $r^2 e^{-r/3a}$ | $\sqrt{\frac{15}{4}}\sin\theta\cos\theta$ | $\frac{e^{\pm i\phi}}{\sqrt{2\pi}}$ | $3\,D_{5/2}$ | 6 |
| 3, 2, ±2 | $\frac{8}{5 \cdot 3^9 \cdot a^7}$ | $r^2 e^{-r/3a}$ | $\frac{\sqrt{15}}{4}\sin\theta^2$ | $\frac{e^{\pm 2i\phi}}{\sqrt{2\pi}}$ | | |
| 7, 3, 0 | $C_r^2 = \left(\frac{2}{7a}\right)^9 \left(\frac{1}{84 \cdot 10!}\right)$ | | $\sqrt{\frac{7}{8}}(5\cos\theta^3 - 3\cos\theta)$ | $\frac{1}{\sqrt{2\pi}}$ | | |
| 7, 3, ±1 | | $\left\{ R(r) = r^3 e^{-2r/7a}\left[-\left(\frac{2}{7a}\right)^3 r^3 + \cdots \right.\right.$ | $\sqrt{\frac{21}{32}}\sin\theta(5\cos\theta^2 - 1)$ | $\frac{e^{\pm i\phi}}{\sqrt{2\pi}}$ | $7\,F_{5/2}$ | 6 |
| 7, 3, ±2 | | $\left. 30\left(\frac{2}{7a}\right)^2 r^2 - 270\left(\frac{2}{7a}\right)r + 720\right] \right\}$ | $\sqrt{\frac{105}{16}}\sin\theta^2\cos\theta$ | $\frac{e^{\pm 2i\phi}}{\sqrt{2\pi}}$ | $7\,F_{7/2}$ | 8 |
| 7, 3, ±3 | | Basis of R(r)[31] | $\sqrt{\frac{70}{64}}\sin\theta^3$ | $\frac{e^{\pm 3i\phi}}{\sqrt{2\pi}}$ | | |

The *intrinsic* component of the kinetic energy ($KE_r$) was obtained in a similar way, noting that both angular integrations again normalize to unity,

$$KE_r = -\frac{\hbar^2}{2m}C_r^2 \int_o^\infty R_{n,l}(r)\left[\frac{1}{r^2}\frac{\partial}{\partial r}\left(r^2 \frac{\partial R_{n,l}(r)}{\partial r}\right)\right]r^2\,dr. \quad (5)$$

The θ, φ operators in Eq. (3) have $r^2\sin\theta$ and $r^2\sin\theta^2$ denominator terms; these were separated and grouped with the appropriate integration variables in Eqs. (6) and (7) together with parts from the element of volume $r^2\sin\theta d\theta d\phi dr$, giving the two *dynamic* kinetic energies $KE_\theta$ (when $l \neq 0$) and $KE_\varphi$ (when $m_l \neq 0$),

$$KE_\theta = -\frac{\hbar^2}{2m}C_r^2 \int_0^\infty \frac{R_{n,l}(r)^2}{r^2}r^2\,dr \times \int_o^\pi \left[\frac{\Theta(\theta)}{\sin\theta}\frac{\partial}{\partial\theta}\left(\sin\theta\frac{\partial\Theta(\theta)}{\partial\theta}\right)\right]\sin\theta d\theta, \quad (6)$$

$$KE_\phi = -\frac{\hbar^2}{2m}C_r^2 \int_o^\infty \frac{R_{n,\ell}(r)^2}{r^2}r^2\,dr \times \int_o^\pi \frac{\Theta(\theta)^2}{\sin\theta^2}\sin\theta d\theta$$
$$\times \frac{1}{2\pi}\int_o^{2\pi} e^{-im_\ell\phi}\frac{\partial^2 e^{im_\ell\phi}}{\partial\phi^2}\,d\phi. \quad (7)$$

Table II lists all energy components from Eqs. (4)–(7). All analyses are archived[32] but the 3, 2, 2 state analyses are also given in the Appendix.

Significantly, all *total* kinetic and *total* kinetic and potential energy values computed by integration over the wave function fields were the same as those obtained from the separation of variables. The energies in Table II derived from the energy operators acting on the fields show them to be real and time independent and that *charge and mass must be distributed as* $\Psi^*\Psi$. The constraints on energies provided valuable computational checksums,

$$E_{total} = \underbrace{-\frac{me^4}{n^2\hbar^2}}_{\text{Potential}=-\frac{2E_1}{n^2}} + \underbrace{(KE_r + KE_\theta + KE_\phi)}_{\text{Kinetic}=\frac{me^4}{2n^2\hbar^2}=\frac{E_1}{n^2}} = \frac{-E_1}{n^2}. \quad (8)$$

These results conformed with the virial theorem, where the kinetic energy is half of the magnitude of the potential energy.[33]

With spin S included, all states with the same n, l fall into either of two groups according to $J = \ell \pm 1/2$, while the range of total momentum projections $J_z$ runs from J to −J. Each $|J, J_z\rangle$ state is a sum of two l, $m_i$ and S, $S_{zi}$ product states, as shown in Eq. (9) (and single states whenever $J = \ell + 1/2, J_z = \pm J$),

$$|J, J_z\rangle = |\ell, m_1\rangle |S, S_{z1}\rangle + |\ell, m_2\rangle |S, S_{z2}\rangle, \text{ where } J_z = m_i + S_{zi}. \quad (9)$$

**TABLE II.** Total kinetic energies $E_n = E_1/n^2$, with $E_1 = m e^4/2\hbar^2$, and potential energies $V_n = -2 E_1/n^2$ determined by integrating Eqs. (4)-(7). Note ($KE_\theta + KE_\varphi$) highlights spinning field energy independence of $m_l$. Energy computations for $n = 1-3$ are archived,[32] as are those for $n = 7$, $l = 3$.[31]

| State | Total KE = ($KE_r + KE_\theta + KE_\varphi$) | | | | | | Total |
|---|---|---|---|---|---|---|---|
| n, l, m | $KE_r$ Intrinsic | $KE_\theta$ | $KE_\varphi$ | ($KE_\theta + KE_\varphi$) Dynamic | KE | Potential $V_n$ | $V_n + $ KE |
| **1**, 0, 0 | $E_1$ | 0 | 0 | 0 | $E_1$ | $-2E_1$ | $-E_1$ |
| **2**, 0, 0 | $E_2$ | 0 | 0 | 0 | $E_2$ | $-2E_2$ | $-E_2$ |
| **3**, 0, 0 | $E_3$ | 0 | 0 | 0 | $E_3$ | $-2E_3$ | $-E_3$ |
| **2**, 1, 0 | $\frac{E_2}{3}$ | $\frac{2}{3}E_2$ | $E_2$ | $\frac{2}{3}E_2$ | $E_2$ | $-2E_2$ | $-E_2$ |
| **2**, 1, ±1 | $\frac{E_2}{3}$ | $\frac{1}{6}E_2 6$ | $\frac{E_2}{2}$ | $\frac{2}{3}E_2$ | $E_2$ | $-2E_2$ | $-E_2$ |
| **3**, 1, 0 | $\frac{5}{9}E_3$ | $\frac{4}{9}E_3$ | o | $\frac{4}{9}E_3$ | $E_3$ | $-2E_3$ | $-E_3$ |
| **3**, 1, ±1 | $\frac{5}{9}E_3$ | $\frac{4}{9}E_3$ | $\frac{3}{9}E_3$ | $\frac{4}{9}E_3$ | $E_3$ | $-2E_3$ | $-E_3$ |
| **3**, 2, 0 | $\frac{3}{15}E_3$ | $\frac{12}{15}E_3$ | o | $\frac{12}{15}E_3$ | $E_3$ | $-2E_3$ | $-E_3$ |
| **3**, 2, ∓1 | $\frac{3}{15}E_3$ | $\frac{7}{15}E_3$ | $\frac{5}{15}E_3$ | $\frac{12}{15}E_3$ | $E_3$ | $-2E_3$ | $-E_3$ |
| **3**, 2, ±2 | $\frac{3}{15}E_3$ | $\frac{2}{15}E_3$ | $\frac{10}{15}E_3$ | $\frac{12}{15}E_3$ | $E_3$ | $-2E_3$ | $-E_3$ |
| **7**, 3, 0 | $\frac{25}{49}E_7$ | $\frac{24}{49}E_7$ | o | $\frac{24}{49}E_7$ | $E_7$ | $-2E_7$ | $-E_7$ |
| **7**, 3, ±1 | $\frac{25}{49}E_7$ | $\frac{17}{49}E_7$ | $\frac{7}{49}E_7$ | $\frac{24}{49}E_7$ | $E_7$ | $-2E_7$ | $-E_7$ |
| **7**, 3, ±2 | $\frac{25}{49}E_7$ | $\frac{10}{49}E_7$ | $\frac{14}{49}E_7$ | $\frac{24}{49}E_7$ | $E_7$ | $-2E_7$ | $-E_7$ |
| **7**, 3, ±3 | $\frac{25}{49}E_7$ | $\frac{3}{49}E_7$ | $\frac{21}{49}E_7$ | $\frac{24}{49}E_7$ | $E_7$ | $-2E_7$ | $-E_7$ |

This mixing of $Y_{l,m}$ basis spatial states does change them, so all the results in Table II are still valid with spin. This can be verified with the comparisons of spectral data with analyses that include spin orbit.[26]

If the electron were a particle, it would have to cover a range of radial positions to satisfy the $\Psi^*\Psi$ distribution, and its potential and kinetic energies would be constantly varying. However, the energy analyses in Table II do not support such a picture; all energies are *not time-averaged*; they are *constant. This is only possible with an electron field representation where the field is everywhere.*

The difficulties of supporting an electron particle representation are more evident with higher quantum states. Figure 1 is a section plot at z = 0 of the H atom $|\psi_{7,3,3}|^2$ state. With spin $|J = 7/2$, $J_z = 7/2\rangle = |3, 3\rangle |1/2, 1/2\rangle$. There are three field intensity lobe pairs along the φ axis, and each of these has 4 lobes moving out along the radial axis. It would be absurd to expect a particle electron to cross zero intensity nodes, change velocity to satisfy the probability function, and *zigzag* in moving between lobes on different axes. The 7, 3, 3 *electron* state's field subdivides into 4 × 6 field components. Similar field patterns can be seen in fiber optic laser modes.[34] In transitioning to another state, all field components appear to "know" they *must* come together in the new state. Roger Penrose has commented similarly on apparent *hidden* photon subdivision when interacting with beam splitters.[35]

### 1. Additional considerations

There are no radial currents $\mathbf{j}_r$ in the H atom,[23(b)] where

$$\mathbf{j}_r = \frac{i\hbar}{2\,m}\left(\Psi \frac{\partial}{\partial r}\Psi^* - \Psi^* \frac{\partial}{\partial r}\Psi\right) = 0. \qquad (11)$$

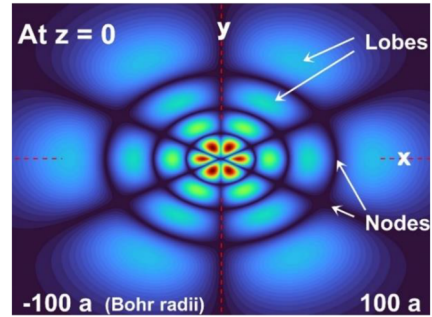

**FIG. 1.** H atom $|\psi_{7,3,3}|^2$ section looking down z-axis.

The 1S state's field *expansion* during ionization is accompanied by *field energy decreases* while *mass increases* by δm until $\delta m\, c^2$ is exactly 13.6 eV.[36] There is no basis for particles to exhibit such energy-to-mass changes.

There is a lack of consistency when the term *matter fields* is correctly used for free electrons, but this interpretation is not acknowledged for those very same electrons in atoms when treated as bound "particles."

### B. Angular momenta n = 2, *l* = 1

An explicit total angular momentum analysis of the $Y_{1,0}$ state is easy on account of its rotational symmetry about the z-axis, but the $Y_{1,±1}$ states, lacking this symmetry, are difficult. With spins $\chi_s$ necessarily included, analysis requires looking at the total momentum **J** states like that given in the following equation:

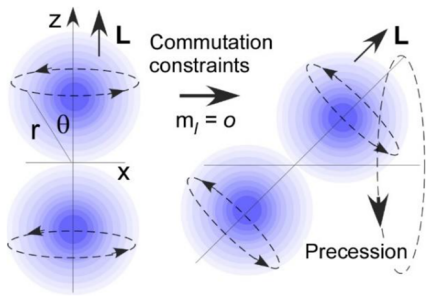

**FIG. 2.** 2 1 0 Orbiting field analysis.

$$|J = 1/2, J_z = 1/2\rangle = \sqrt{\frac{1}{3}} Y_{1,0} \chi_+ - \sqrt{\frac{2}{3}} Y_{1,1} \chi_-. \quad (12)$$

All $Y_{1,m}$ basis states for a given n, l have the same total kinetic energy (KE) and the same angular momentum, but in the mixed representation these contributions are proportionally distributed as 1/3 and 2/3. Just like the small spin orbit energy offsets are apportioned between the component states,[26] analyses can proceed based only on the $Y_{1,0}$ basis state, but with the understanding that the results are really to be shared between the two $Y_{l,m}$ components of $|J, J_z\rangle$.

The field velocity $v_{r,\theta}$ at any point r, θ for this state is illustrated by arrows in Fig. 2. The period T can be determined by summing all field contributions δ**L** at r, θ to the total $|\mathbf{L}| = (l(l+1))^{1/2}\hbar$,

$$|\delta \mathbf{L}| = |\mathbf{v}_{r,\theta} \times \mathbf{r} \sin\theta|\delta m, \quad (13)$$

where

$$v_{r,\theta} = 2\pi r\sin\theta/T, \quad \delta m = m|\Psi|^2 r^2 \sin\theta dr d\theta d\phi, \quad (14)$$

and

$$|\mathbf{L}| = \left(\frac{3m}{2T}\right)\int_o^\infty C_r^2 r^6 e^{-r/a} dr \times \int_o^\pi \cos\theta^2 \sin\theta^3 d\theta \times \int_o^{2\pi} d\phi = 2^{1/2}\hbar, \quad (15)$$

$$T = 24 \ m\pi a^2 /\sqrt{2}\ \hbar. \quad (16)$$

The field energy contributions are then

$$\delta E = \frac{1}{2}\left(\frac{2\pi r\sin\theta}{T}\right)^2 m|\Psi|^2 \ dV, \quad (17)$$

$$E = \left(\frac{3\pi m}{2T^2}\right)\int_o^\infty C_r^2 r^6 e^{-r/a} dr \times \int_o^\pi \cos\theta^2 \sin\theta^3 d\theta \times \int_o^{2\pi} d\phi. \quad (18)$$

Substituting for T, $a = \hbar^2/me^2$, $E_1 = me^4/2\hbar^2$ gave a dynamic energy of the spinning field of $2E_1/3n^2$. This was the same energy as that derived from the operator formalism, but as mentioned earlier, this energy is actually shared between the two spatial states $Y_{1,0}$ and $Y_{1,1}$.

The field spinning frequency $f_{field\_2p} \sim 8 \times 10^{14}$ Hz can be compared with a *particle orbiting* model with $e^2/r^2 = m r \omega^2$ gives $f_{particle\_2p} \sim 5 \times 10^{14}$ Hz, where $\langle r^3 \rangle = 210\ a^3$ for 2p states. These frequencies are not far apart.

A precession rate of $\sim 10^8$ Hz has been determined.[37] This has a negligible energy impact. At a distance of 3 Bohr radii and with sin θ at a maximum, the field velocity is ∼0.3% c. This justified the first order nonrelativistic analysis.

### III. OTHER REQUIREMENTS BASED ON FIELDS

Many other properties can *only* be explained with fields; a few are listed as follows:

(1) The well-known "Particle in Box" problem should properly be described as the "Particle Field in Box." If the box walls are increased by integer multiples n, the lowest mode energy will decrease in the same way as n increases in the H atom. Most QM books neglect to explicitly define the energies in this problem as *field* energies. Similarly, in some metals, it is outer electron *delocalization* that partly leads to their bonding strength.[38]

(2) The *time independent* and *continuous spherical* distribution of field charge in *closed* n l subshells in atoms is what enables these fields to *interpenetrate each other smoothly*, resulting in very highly reproducible spectral properties.

(3) Tunneling can be physically explained with a *field* representation because boundary conditions just inside the barrier demand *an increase in the field's second derivative and, therefore, an* increased *energy density*.

(4) Because fermion fields *overlap,* they can better "feel" each other over large separations and thereby abide by the exclusion principle. This cannot be so easily appreciated if fermions were *isolated particles*.

The properties of scattering and diffraction can be added to this list.

### IV. DISCUSSION

While state collapse was not a part of this study, it needs to be said that it has often been misinterpreted as the *probability of finding a particle on measurement*. This is a vestige of what Max Born suggested in the early days of quantum mechanics before the concept of quantum fields came into existence.[39] Collapsed states are sometimes described as "particle-like" simply because the collapse interaction takes place in a very small region. The problem is exemplified by quoting from well-known physicist Albert Messiah[40] on matter waves: "*The conditions for the validity of Classical Mechanics are fulfilled when the wave maintains in the course of time a sufficiently small extension so that it may be approximated by a point, and one may attribute a precise motion to the particle.*" However, even a very small size does not justify a field/wave transformation into a particle! This perception of small size stems from the fact that the collapse interaction can only involve the excitation of, *at most a few atoms*. Hence, the illusion of collapse to a very small point. The initial state represents a real *field*, and the collapsed state is either a changed *field* state or it is completely absorbed by exciting a field state(s) in the detecting medium.

Does it matter if we keep using the term "particle" with no qualifiers if Schrödinger's equation works just fine? It does matter, as two physicists[41] explained: "*If you fail to address philosophical issues head-on, they do not go away, they come around and bite* you on the rear." It is concerning when many physicists still report so many differences of opinion on fundamental QM interpretations (Q3).[42] Misrepresenting electrons or taking the attitude "Shut Up and Calculate" is detrimental to progress in physics. Common agreement on a fundamental matter like fields is vital.

It is understandable that in 1927 it must have been difficult to accept all matter as all being nebulous fields, and so the Copenhagen Interpretation was enshrined. However, a century later, it seems clear that the concepts of real particles need to be dropped. Perhaps Carlo Rovelli's[43] remarks may be helpful: "*Scientific thinking is not made up of certainties, it is thinking constantly in motion, the power of which is to always question everything and begin over again, to be fearless in subverting the order of the world in search of a more efficient one.*"

## V. CONCLUSIONS

This study supports the thesis that Schrödinger's equation *involves fields* with charge and mass distributed as the normalized intensity function. Interactions, detections or excitations can be characterized as field-to-field transitions occurring with a probability proportional to the field intensity. All quantum descriptors should flow from the fundamental basis of fields.

Schrödinger's equation *only appears* to work as a particle equation, but probing into the matter revealed several inconsistencies while some properties simply cannot be explained with particles.

## ACKNOWLEDGMENTS

This work was prompted by the need to resolve the different points of view between myself and my colleague, Dr. Alex Malozemoff, while we were writing a paper entitled *Quantum Misuse in Psychic Phenomena*, published in the Journal of Near Death Studies. I am deeply grateful to Alex for providing the motivation needed to engage in this study. I acknowledge Professor Zdzislaw Musielak for his helpful suggestions and Rand Hall from our library for his support providing reference materials.

## AUTHOR DECLARATIONS
### Conflict of Interest

The author has no conflicts to disclose.

### Author Contributions

**Jacek Mroczkowski**: Conceptualization (equal); Data curation (equal); Formal analysis (equal); Investigation (equal); Methodology (equal); Software (equal); Validation (equal); Visualization (equal); Writing – original draft (equal); Writing – review & editing (equal).

## DATA AVAILABILITY

The data that support the findings of this study are openly available in Zenodo.[26,31,32,36,37]

## APPENDIX: n = 3, *l* = 2, m = ∓2 EXAMPLE OF ENERGY COMPUTATIONS IN TABLE II

$$KE_r = -\frac{\hbar^2}{2m}\int_o^\infty C_r^2\left(6r^2 - \frac{2r^3}{a} + \frac{r^4}{9a^2}\right)r^2 e^{-2r/3a}\,dr$$

$$= -\frac{\hbar^2}{2m}\frac{8}{5\cdot 3^9 a^7}\left(\frac{6\cdot 4!3^5}{2^5} - \frac{2\cdot 5!3^6}{2^6} + \frac{6!3^7}{9\cdot 2^7}\right)a^5. \quad (A1)$$

$$= \frac{3E_1}{15n^2}\left(\text{where } a = \frac{\hbar^2}{me^2}, E_1 = \frac{me^4}{2\hbar^2}\right). \quad (A2)$$

$$-\frac{\hbar^2}{2m}\times \text{Radial integral} = \textbf{RI term} = -\frac{\hbar^2}{2m}\int_o^\infty C_r^2 r^4 e^{-2r/3a}\,dr$$

$$= -\frac{2}{15n^2}\frac{me^4}{2\hbar^2}. \quad (A3)$$

$$\textbf{Polar term} = \left(\frac{15}{16}\right)\int_o^\pi \sin\theta^2 \left[\frac{1}{r^2\sin\theta}\frac{\partial}{\partial\theta}\left(\sin\theta\frac{\partial}{\partial\theta}(\sin\theta^2)\right)\right]r^2$$

$$\times \sin\theta\,d\theta \times \frac{1}{2\pi}\int_o^{2\pi}d\phi = -1. \quad (A4)$$

$$\textbf{KE}_\theta = \textbf{RI term}\times \textbf{Polar term} = \frac{2E_1}{15n^2}. \quad (A5)$$

$$\textbf{Az term} = \left(\frac{15}{16}\right)\int_o^\pi \frac{\sin\theta^5}{\sin\theta^2}d\theta\frac{1}{2\pi}\int_o^{2\pi}\frac{\partial^2}{\partial\phi^2}e^{\pm 2i\phi}\,d\phi$$

$$= \left(\frac{15}{16\pi}\right)\int_o^\pi \sin\theta^3 d\theta(\pm 2i)^2\int_o^{2\pi}d\phi = -5, \quad (A6)$$

$$\textbf{KE}_\phi = \textbf{RI term}\times \textbf{Az term} = \frac{10E_1}{15n^2}, \quad (A7)$$

$$\text{Potential energy } \textbf{V} = -e^2 C_r^2\int_o^\infty \frac{1}{r}r^4 e^{-2r/3a}r^2\,dr$$

$$= -e^2\left(\frac{8}{5\cdot 3^9 a^7}\right)\left(\frac{5!3^6 a^6}{2^6}\right) = -\frac{2E_1}{n^2}, \quad (A8)$$

$$\text{Total energy} = \textbf{KE}_r + (\textbf{KE}_\theta + \textbf{KE}_\phi) + \textbf{V} = -\frac{E_1}{n^2}. \quad (A9)$$

## REFERENCES

[1] G. Auletta *et al.*, *Quantum Mechanics* (Cambridge University Press, 2009), p. 406.
[2] M. Kaku, *Physics of the Impossible* (Doubleday, 2008), p. 59.
[3] J. Basdevant and J. Dalibard, *Quantum Mechanics* (Springer, 2005), p. 240.
[4] D. Griffiths, *Introduction to Quantum Mechanics* (Cambridge Press, 1995), p. 11.
[5] D. Griffiths and D. Schroeter, *Introduction to Quantum Mechanics* (Prentice-Hall, 2018), p. 166.
[6] J. Townsend, *Quantum Physics* (University Science Books, 2010), p. 64.
[7] R. Omnes, *Understanding Quantum Mechanics* (Princeton University Press, 2020), p. 49.
[8] C. Rovelli, *Reality Is Not What it Seems* (River Head Books, 2017), p. 126.
[9] F. Wilczek, *Fundamentals: Ten Keys to Reality* (Penguin Press, 2021), p. 102.

[10] S. Weinberg, *Facing Up: Science and its Cultural Adversaries* (Harvard Press, 2001), p. 109.
[11] A. Hobson, "There are no particles, there are only fields," Am. J. Phys. **81**, 211 (2013).
[12] A. Hobson, *Fields and Their Quanta: Making Sense of Quantum Foundations* (Springer Nature, 2024), p. 79.
[13] (a) W. Moore, *Schrodinger Life and Thought* (Bath Press, 1989), p. 219; (b) W. Moore, *Schrodinger Life and Thought* (Bath Press, 1989), p. 221; (c) W. Moore, *Schrodinger Life and Thought* (Bath Press, 1989), p. 240.
[14] (a) J. Gribben, Jr., *Erwin Schrodinger and the Quantum Revolution* (Wiley, 2012), p. 228; (b) J. Gribben, Jr., *Erwin Schrodinger and the Quantum Revolution* (Wiley, 2012), p. 152; (c) J. Gribben, Jr., *Erwin Schrodinger and the Quantum Revolution* (Wiley, 2012), pp. 146–147; (d) J. Gribben, Jr., *Erwin Schrodinger and the Quantum Revolution* (Wiley, 2012), p. 154.
[15] H. Kragh, *Dirac: A Scientific Biography* (Cambridge University Press, 1990), p. 47.
[16] E. Inonu and E. Wigner, "Representations of the Galilei group," Nuovo Cimento **9**, 705 (1952); *The Collected Works of Eugene Paul Wigner*, edited by A. S. Wightman (Springer, 1993), Vol. 1, p. 459.
[17] M. Born, "The interpretation of quantum mechanics," Br. J. Philos. Sci. **4**(14), 95 (1953).
[18] B. d'Espagnat, *Veiled Reality* (Addison-Wesley, 1995), p. 14 or West View—2005.
[19] R. Omnes, *Understanding Quantum Mechanics* (Princeton Press, 1999), p. 49.
[20] J. Bell, "Against measurement," Phys. World **3**(8), 39 (1990).
[21] A. Hobson, *Tales of the Quantum* (Oxford University Press, 2017), p. 93.
[22] L. Smolin, *The Trouble with Physics* (Houghton M, 2006), pp. 264–265, 294–295, 300, 330.
[23] (a) W. Greiner, *Quantum Mechanics*, 4th ed. (Springer, 2001), p. 145; (b) W. Greiner, *Quantum Mechanics*, 4th ed. (Springer, 2001), p. 228.
[24] T. Norsen, *Foundations of Quantum Mechanics* (Springer, 2017), p. 120.
[25] R. Feynman *et al.*, in *Lectures on Physics*, edited by N. Millen (Basic Books, 2010–2011), Vol. 3, p. 21–4.
[26] J. Mroczkowski (2025). "Spin orbit and coupling coeffs," Zenodo. https://zenodo.org/records/14767002
[27] R. Shankar, *Principles of Quantum Mechanics* (Springer, 1980), p. 581.
[28] D. Anderson and K. Krzyzanowska, "A gauge field theory of coherent matter waves," AVS Quantum Sci. **5**, 034403-8 (2023).
[29] R. Dicke and J. Wittke, *Introduction to Quantum Mechanics* (Addison-Wesley, 1960), p. 143.
[30] L. Schiff, *Quantum Mechanics*, 3rd ed. (McGraw-Hill, 1968), p. 77.
[31] J. Mroczkowski (2024). "H73_m_state_analysis (version 1)," Zenodo. https://zenodo.org/records/13730969
[32] J. Mroczkowski (2024). "Table II computations," Zenodo. https://zenodo.org/records/13732020
[33] I. Prigogine and S. Rice, *Advances in Chemical Physics* (Intertec Publishing, 1985), Vol. 61, p. 245.
[34] G. Yu *et al.*, "Design of miniaturization resonant cavities using metamaterial," Open Phys. **10**(1), 140 (2012).
[35] R. Penrose, *The Road to Reality* (Alfred A. Knopf, New York, 2005), p. 514.
[36] J. Mroczkowski (2024). "Energy," Zenodo. https://zenodo.org/records/13732039
[37] J. Mroczkowski (2024). "Electron precession," Zenodo. https://zenodo.org/records/13860926
[38] J. Christman, *Fundamentals of Solid State Physics* (John Wiley & Sons, Inc., 1988), p. 132.
[39] M. Born, *My Life: Recollections of a Nobel Laureate* (Scribner, 1978), p. 232.
[40] A. Messiah, *Quantum Mechanics*, 2nd ed. (Dover, 2014), p. 49.
[41] R. Crease and A. Goldhaber, *The Quantum Moment* (W. W. Norton & Company, 2014), p. 260.
[42] M. Schlosshauer, *Elegance and Enigma* (Springer, 2011).
[43] C. Rovelli, *Helgoland* (Riverhead Books, New York, 2021), p. 73.